\pdfoutput=1

\documentclass[grl]{agu2001}

\usepackage{graphicx}
\usepackage{amssymb}
\usepackage[usenames]{color}
\usepackage{lineno}

\authorrunninghead{ANDREOTTI ET AL.}

\titlerunninghead{Measurements of the aeolian saturation length}

\authoraddr{PMMH, UMR 7636 CNRS--ESPCI--Univ.~P6-P7, 10 Rue Vauquelin, 75231 Paris Cedex 05, France.\\
(\texttt{andreotti@pmmh.espci.fr}, \texttt{claudin@pmmh.espci.fr})}

\authoraddr{IUSTI, UMR 6595 CNRS--Univ.~Provence, 5 rue Enrico Fermi, 13453 Marseille Cedex 13, France.\\
(\texttt{Olivier.Pouliquen@univ-provence.fr})}

\begin{document}

\title{
Measurements of the aeolian sand transport saturation length}

\author{
B. Andreotti$^\sharp$, P. Claudin$^\sharp$ and O. Pouliquen$^\bigstar$}

\affil{
$^\sharp$Laboratoire de Physique et M\'ecanique des Milieux H\'et\'erog\`enes (PMMH),\\
UMR 7636 CNRS--ESPCI--Univ.~P6-P7, 10 Rue Vauquelin, 75231 Paris Cedex 05, France.\\
$^\bigstar$Institut Universitaire des Syst\`emes Thermiques Industriels (IUSTI),\\
UMR 6595 CNRS--Univ.~Provence, 5 rue Enrico Fermi, 13453 Marseille Cedex 13, France.
}

\begin{abstract}
The wavelength at which a dune pattern emerges from a flat sand bed is controlled by the sediment transport saturation length, which is the length needed for the sand flux to adapt to a change of wind strength. The influence of the wind shear velocity on this saturation length and on the subsequent dune initial wavelength has remained controversial. In this letter, we present direct measurements of the saturation length performed in a wind tunnel experiment. In complement, initial dune wavelengths are measured under different wind conditions --~in particular after storms. Using the linear stability analysis of dune formation, it is then possible to deduce the saturation length from field data. Both direct and indirect measurements agree that the saturation length is almost independent of the wind strength. This demonstrates that, in contrast with erosion, grain inertia is the dominant dynamical mechanism limiting sediment transport saturation.
\end{abstract}

\begin{article}

\section{Introduction}
\label{intro}

Understanding how wind transports sand is important in many geomorphological problems including dunes dynamics. It is well known that for a given wind velocity $u_*$ greater than a threshold velocity $u_{\rm th}$, the sand flux $q$ saturates to a steady value $q_{\rm sat}$, which results from a subtile interaction between grains motion and wind velocity. Although the origin of the saturation flux is now well understood, the transient regime and the way $q$ adjusts to $q_{\rm sat}$ still remains controversial. If a change of velocity takes place, the flux does not adjust instantaneously to its equilibrium value, and a phase lag exists between $q$ and velocity $u_*$. Whatever the dynamical mechanisms responsible for this lag, it has been proposed to encode it into a single length $L_{\rm sat}$, called the saturation length. By definition,  $L_{\rm sat}$ is the length over which the flux $q$ relaxes to its equilibrium value $q_{\rm sat}$. Around the saturated regime, this relaxation can be modeled by a first order equation:

\begin{equation}
L_{\rm sat} \frac{dq}{dx}=q_{\rm sat}-q.
\label{EquaSaturation}
\end{equation}

DIfferent physical mechanisms can be responsible for this saturation length  e.g., the grain hop length [\textit{Charru 2006}], the length needed to accelerate new grains (called the drag length) [\textit{Andreotti et al. 2002}, \textit{Hersen  et al. 2002}, \textit{Andreotti and Claudin 2007}], the length needed to expel new grains from the sand bed [\textit{Sauermann  et al. 2001}; \textit{Parteli  et al. 2007}], the length needed for the negative feedback of transport on the wind to take place [\textit{Andreotti, 2004}]. Transport saturation is limited by the \emph{slowest} of these processes.

Understanding the physical origin of sand flux saturation is crucial to better describe the formation of dunes. Dunes result from the interaction between the wind and their shape  [\textit{Bagnold 1941}, \textit{Lancaster et al. 1996}, \textit{Wiggs et al. 1996}, \textit{McKenna-Neuman et al. 1997}, \textit{Andreotti et al. 2002}]. Recent models have shown that the initial formation of dunes is controlled by the balance between three different mechanisms. The first one is the phase-lag between the dune elevation profile $h$ and the basal shear velocity $u_*$ [\textit{Jackson and Hunt 1975}]. This purely hydrodynamical mechanism, related to fluid inertia and dissipation, is the motor o the instability and tends to destabilise the bed [\textit{Richards 1980}, \textit{Andreotti et al. 2002}, \textit{Elbelrhiti et al. 2005}, \textit{Fourri\`ere et al. 2008}]. On the other hand, two stabilizing effects exist. The dynamics of sediment transport tends to stabilise short wavelengths, due to  the space-lag between sand flux $q$ and shear velocity $u_*$.  Gravity has a further stabilizing effect, related to the dependence of the threshold shear velocity $u_{\rm th}$ on the local slope [\textit{Rasmussen et al. 1996}].  A stability analysis taken into account the three ingredients, [\textit{Andreotti and Claudin 2007}] shows that the most unstable wavelength is proportional to $L_{\rm sat}$ and decreases with $u_*/u_{\rm th}$. Saturation length then controls the initial size of dunes. 

In this letter, we aim to investigate the physical origin of the sand transport saturation length by studying its dependence on the wind strength. Two methods are used. A direct one by measuring the saturation of the sand flux in a wind tunnel, and an indirect method based on field measurements of the instability of sand bed. 
\begin{figure}[h!]
\centerline{\includegraphics[width=\columnwidth]{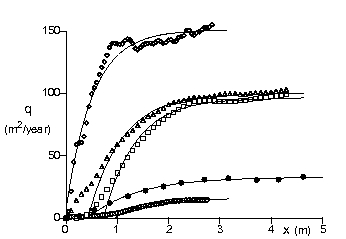}}
\caption{Spatial variation of the sediment flux over a flat sand bed, measured in a wind tunnel for different wind speeds: $u_*=0.25$m/s ($\circ$), $u_*=0.33$m/s ($\square$), $u_*=0.34$m/s ($\triangle$) and $u_*=0.42$m/s ($\diamond$).  In experiments ($\triangle$) and ($\diamond$) a small input flux was injected to initiate saltation.  Solid lines: exponential fit (see text). Field measurement \textit{Elbelrhiti et al. 2005} for $d=185\mu$m and $u_*\sim 1.3 u_{\rm th}$ are also plotted ($\bullet$).}
\label{qofx}
\end{figure}
%

\section{Wind tunnel measurements}
\label{wind_tunnel}
\begin{figure*}[t!]
\centerline{\includegraphics[width=\textwidth]{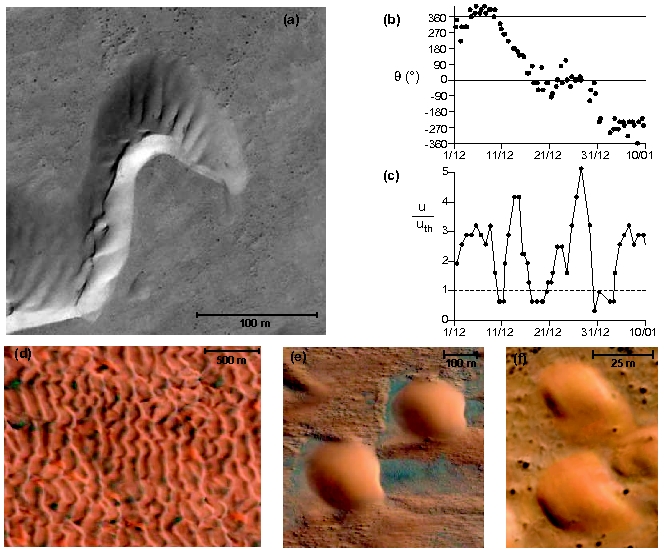}}
\caption{(a) Barchan instability in the Atlantic Sahara due to a storm characterised by a strong dry wind coming from inland (Chergui). Aerial photograph taken on january the 5$^{\rm th}$, 2005 --~credits DigitalGlobe. (b) Wind direction measured in Tan-Tan between december the 1$^{\rm st}$, 2004, and january the 10$^{\rm th}$, 2005 --~regular trade winds are along $\theta\simeq23^{\rm o}$. (b) Corresponding values of the basal shear velocity $u_*$. (d) Aerial picture of a transverse dune field in Northwestern Ar-Rub-Alkhali (S. Arabia). (e) $220$m long dunes in Northwestern Ar-Rub-Alkhali (S. Arabia) (typical local wind: $u_*\sim 1.2 u_{\rm th}$). (f) $30$m long dunes in the Atlantic Sahara (typical local wind: $u_*\sim 1.8 u_{\rm th}$).}
\label{FieldWavelength}
\end{figure*}

We first report direct measurements of $L_{\rm sat}$ obtained from controlled experiments. They have been performed in the wind tunnel of the \textsc{Cemagref} in Grenoble ($1$m wide, $0.5$m high, and $4.5$m long), using sand grains of diameter $d=120\pm40\mu$m from the Hostun quarry [\textit{Cierco et al. 2008}]. An initial $3$cm thick sand bed is prepared and flattened with a moving bar. This thickness is gently matched to the wind tunnel rough bottom over the first ten centimetres. The bed elevation profile $h(x,t)$ is measured at regular time interval with a vertical resolution of $500\mu$m, turning off the flow. Using the conservation of mass, the sediment flux $q$ at position $x$ is simply deduced from the erosion data as:
\begin{equation}
q(x)=-\int_0^x \!\! \partial_t h(\xi) \,d\xi
\end{equation}
The curves $q(x)$ obtained for different wind strengths are depicted in figure~\ref{qofx}. In all the cases, the sediment transport increases downstream, starting from a null or small input flux, and further saturates to a value $q_{\rm sat}$, which increases with the wind strength. The evolution of $q$ can be divided in two phases. A first exponential increase followed by a relaxation phase toward the saturation. The initial phase is linked to ejection of grains, each saltating grain ejects few other grains when it collides the bed. This results into an exponential increase of the flux. This process is noticeable for weak wind, and when the flux is low [\textit{Andreotti and Claudin 2007}]. This regime is a priori not taken into account in  eq. 1, which aims to describe the relaxation close to the saturation level. To determine the saturation length, we have thus analysed the zone where the flux is larger than half its saturated value $q_{\rm sat}$.  The solid lines in figure~\ref{qofx} show the best fit by an exponential law of the form $q_{\rm sat}[1 - e^{-(x-x_0)/L_{\rm sat}}]$, which is the solution of equation~(\ref{EquaSaturation}). As expected, $q_{\rm sat}$ iincreases with wind strength. In contrast, the fitted values of $L_{\rm sat}$, on the order of $0.7$m, are remarkably insensitive to those of $u_*$, as well as to the entrance conditions (Fig.~\ref{Lsatofu}).
\begin{figure}[t!]
\centerline{\includegraphics[width=\columnwidth]{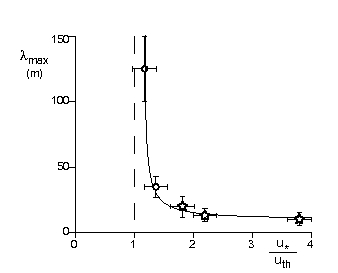}}
\caption{Measured most unstable wavelength $\lambda$ as a function of the rescaled wind shear velocity $u_*/u_{\rm th}$, for a mean grain size $d\simeq185\mu$m. The star symbols correspond to superimposed structures on barchans generated by stormy events. The circles correspond to low amplitude transverse dunes generated by averaged winds. The solid line is a fit by the model derived in \textit{Fourri\`ere et al. 2008} assuming that the saturation length is constant -- it gives $L_{\rm sat}\simeq 68$cm.}
\label{LambdaEolien}
\end{figure}

Figure~\ref{qofx} also displays a flux profile measured in the field by \textit{Elbelrhiti et al. 2005} in the Atlantic Sahara. A $20$m long sand bed composed by grains of size $d=185\mu$m was prepared with a bulldozer flatten. The erosion was measured after $24$~hours of wind fluctuating around the transport threshold. To determine a characteristic wind strength in such a case, we start from velocity time series $U(t)$ measured at some altitude. As shown by \textit{Ungar and Haff 1987} and \textit{Andreotti 2004}, the saturated flux $q_{\rm sat}$ is proportional to $(U^2-U_{\rm th}^2)\,\mathcal{H}(U-U_{\rm th})$, where $U_{\rm th}$ is the threshold value for transport, and where $\mathcal{H}(x)$ is the so-called Heaviside function [$\mathcal{H}(x>0)=1$ and $\mathcal{H}(x<0)=0$]. We then define the effective velocity as that which would give the same mean flux, if the wind was not fluctuating:
\begin{equation}
\left(\frac{u_*}{u_{\rm th}}\right)^2=1+\left<\left[\left(\frac{U(t)}{U_{\rm th}}\right)^2-1\right]\,\mathcal{H}(U(t)-U_{\rm th})\right>,
\label{averaging}
\end{equation}
where the $\left < ... \right >$ denotes an average over time. Interestingly, the ratio so defined does not depend on the height above the soil at which the anemometer is placed, which justifies that it is also the ratio of the typical shear velocity $u_*$ to its threshold value $u_{\rm th}$. For this particular measurement, we find $u_* \simeq 1.3 u_{\rm th}$. By fitting the curve $q(x)$ in the zone where $q>q_{\rm sat}/2$, we find a saturation length of $1\pm0.2$m, twice smaller than that obtained by fitting the whole profile ($1.7$m). The saturation lengths measured here are also significantly smaller than that ($L_{\rm sat}\sim2.3$m) which can be determined from the measurements of \textit{Bagnold 1941} (Fig.~62, p.~182) with grains of diameter $d=240\mu$m. This is probably due to the fact that the erosion rate was measured by means of spring balances below sections of the tunnel, of the size of the actual saturation length. In contrast, the space resolution of $10$cm used in our wind tunnel measurements allows to reduce the error bars on $L_{\rm sat}$ to $10\%$.

\section{Initial dune wavelength}
\label{dune_wavelength}

The analysis of wavelength at which dunes form from a flat sand bed provides an indirect way to measure the saturation length. This inverse problem is made possible thanks to the increased precision of hydrodynamical calculations giving the turbulent velocity field around obstacles of small amplitude [\textit{Jackson and Hunt 1975}, \textit{Richards} 1980, \textit{Fourri\`ere et al.} 2008]. Given that the instability requires typically $100$m to develop, wind tunnels cannot be used to form dunes under controlled conditions. Fortunately, the surface of large dunes precisely behave as flat areas of sand and are thus submitted to the primary linear instability [\textit{Elbelrhiti et al. 2005}]. The main difficulty is to assess the wind velocity to be associated to the formation of given superimposed bedforms. Figure~\ref{FieldWavelength}(a) shows a situation in which this can be achieved rigourously. Indeed, the periodic superimposed bedforms of wavelength $\lambda$ on the stoss slope of this barchan dune is transverse to the dune itself. They are formed by a strong storm almost perpendicular to the regular trade winds ($\theta=23^{\rm o}$) blowing over the Atlantic Sahara (Fig.~\ref{FieldWavelength}b). This hot and dry anomalous wind, called `Chergui', is due to the motion of the Azores anticyclone over Europe and generates dust jets over the ocean. Figures~\ref{FieldWavelength}(b) and (c) show the two successive storms of this type, both characterised by  a peak of wind velocity and a rotation of the wind direction. We have analyzed several of such events, which occur between $5$ and $10$ times per year and are presumably responsible for a part of the dynamics of barchan fields [\textit{Hersen  et al. 2004}, \textit{Andreotti and Claudin 2007}]. Importantly, we were present in the field during some of these storms and have been able to observe the resulting bed instability. In order to determine the most unstable wavelength, we have averaged the spacing between secondary bed-forms over typically $20$ barchan dunes. The wind velocity is consistently determined on the crest of dunes, taking into account the speed-up factor $\sim1.4$ [\textit{Andreotti et al. 2002}] with respect to the wind strength over the surrounding flat ground (Fig.~\ref{FieldWavelength}c). It is averaged over the period of time during which the wind blows in the direction perpendicular to the new born crests, following equation~(\ref{averaging}). This procedure allows to obtain data (stars in fig.~\ref{LambdaEolien}) far above the transport threshold (from $1.5 u_{\rm th}$ to $4 u_{\rm th}$), since the selected events should be strong enough to destabilise the surface of dunes within few days. In this range of wind velocity, the most unstable wavelength $\lambda_{\rm max}$ is roughly independent of $u_*$.

\begin{figure}[t!]
\centerline{\includegraphics[width=\columnwidth]{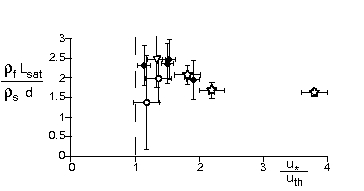}}
\caption{Saturation length $L_{\rm sat}$, rescaled by the inertial length $\rho_s/\rho_f \, d$, as a function of the wind shear velocity $u_*$, rescaled by the threshold $u_{\rm th}$. Direct measurements, performed in a wind tunnel ($\blacklozenge$) and in the field ($\triangle$), are compared to those determined from the initial dune wavelength (storms: $\star$ and slipfaceless dunes ($\circ$)) }
\label{Lsatofu}
\end{figure}

Seeking for situations closer to the threshold of transport, we have investigated the case of dunes that present evanescent avalanche slip faces over long period of time. Figure~\ref{FieldWavelength} shows transverse dunes of this type, in northwestern Ar-Rub-Alkhali (S. Arabia). They are composed of grains of size $d \simeq190 \mu$m [\textit{Abolkhair 1986}], i.e. comparable to those in the Atlantic Sahara. We hypothesise that these small amplitude dunes remain at the wavelength at which a flat sand bed destabilises. They can be considered as submitted to the average wind at the scale of the year, as defined by equation~(\ref{averaging}). One finds a mean wavelength $\lambda=130$m, in Ar-Rub-Alkhali, while the same pattern in the Atlantic Sahara rather presents a smaller wavelength $\lambda=35$m. Analysing time-series of the velocity measured in the airports surrounding these dune fields, it turns out that the characteristic wind velocity is significantly smaller in the former ($u_*\sim 1.2 u_{\rm th}$) than in the later ($u_*\sim 1.4 u_{\rm th}$) location. Because of the assumption made, these two points (circles in Fig.~\ref{LambdaEolien}) are less reliable than those measured after the destabilisation induced by a well identified wind event. Still, two independent observations make us confident that the large wavelength observed in northwestern Ar-Rub-Alkhali is not due to a pattern coarsening process. First, we can not see there any superimposed bedform. Second, in the border of the transverse dune field, one can evidence isolated dunes in an intermediate state between slipfaceless dome dunes and well developed barchans. This is clearly the signature of a dune close to the minimal length of dunes [\textit{Hersen et al. 2002, Kroy  et al. 2002, Parteli et al. 2007, Andreotti and Claudin 2007}]. In Ar-Rub-Alkhali, they can be as large as $220$m, which is $\sim 7$ times larger than in the Atlantic Sahara (compare panels (e) and (f) of Fig.~\ref{FieldWavelength}). This confirms the sharp increase of the most unstable wavelength close to the transport threshold (Fig.~\ref{LambdaEolien}). Based on the comprehensive linear stability analysis derived in \textit{Fourri\`ere et al. 2008}, which includes the effects of size $\lambda/d$ as well as wind $u_*/u_{\rm th}$ ratios, we have solved the inverse problem and converted the data $\lambda_{\rm max}(u_*/u_{\rm th})$ into a relation between $L_{\rm sat}$ and $u_*/u_{\rm th}$. Figure~\ref{Lsatofu} shows that this independent determination of $L_{\rm sat}$ is in fair agreement with the direct one. 
%

\section{Conclusion}
\label{conclu}

In this letter we have presented the first accurate measurements of the sand transport saturation length $L_{\rm sat}$. Both direct and indirect methods consistently show that $L_{\rm sat}$ is mostly independent on the wind strength. This result points to a very simple interpretation related to grain inertia. Each grain dislodged from the bed needs some length to reach its asymptotic velocity. A crude determination of this length can be achieved by solving the equation of motion for the horizontal grain velocity component $v=\frac{dx}{dt}$:
\begin{equation}
\frac{\rho_s \pi d^3}{6}\,\frac{d  v}{dt} = \frac{\rho_f \pi d^2}{8}\,C_d| U- v| ( U- v),
\end{equation}
where $\rho_s$ is the density of the grain, $\rho_f$ the density of the fluid, $U$ the fluid velocity and $C_d$ the turbulent drag coefficient. This equation can be rewritten in a dimensionless form, changing variable from $t$ to the grain position $x$:
\begin{equation}
\tilde{v} \frac{d \tilde{v}}{d \tilde{x}} = |1- \tilde{v}| (1- \tilde{v}), \,\,\,\, {\rm with} \,\,\,\, \tilde{v}=\frac{v}{U} \,\,\,\,\&\,\,\,\, \tilde{x}=\frac{3 \rho_f C_d}{4 \rho_s d } x\,\,\,\,
\end{equation}
showing that there is a single length scale in the problem, which does not depend on the wind velocity $U$. As a consequence, the solution of this equation, $\tilde{x}=\tilde{v}/(1-\tilde{v}) + \ln(1-\tilde{v})$ shows a spatial transient over a length proportional to $\rho_s/\rho_f \, d$. As this is precisely the scaling law followed by $L_{\rm sat}$ [\textit{Claudin and Andreotti 2006}], the saturation transient of the sediment flux is certainly controlled by the grain inertia.

\noindent
\rule[0.1cm]{3cm}{1pt}

We wish to thank F. Naaim and the ETNA group of the \textsc{Cemagref} in Grenoble for the use of their wind tunnel and for their kind help during the experiments. This work has been partially supported by the french ministry of research as well as by the `Institut Universitaire de France'.


\end{article}
\end{document}